\documentclass[11pt,epsf]{article}

\usepackage{graphicx}
\usepackage{amssymb}
\usepackage{amsmath}
\usepackage{bm}

\def\be{\begin{eqnarray}}
\def\ee{\end{eqnarray}}
\def\MeV{\mbox{MeV}}

\def\fm{\mbox{fm}}
\def\roughly#1{\mathrel{\raise.3ex\hbox{$#1$\kern-.75em%
\lower1ex\hbox{$\sim$}}}}

\def\gsim{\roughly>}

\newcommand{\ba}{\begin{eqnarray}}
\newcommand{\ea}{\end{eqnarray}}

\begin{document}

\begin{titlepage}

\hfill {\today }

\begin{center}

\centerline{\Large\bf Dependence of relative abundances of
constituents in dense stellar matter}

\centerline{\Large\bf  on nuclear symmetry energy }

\vspace{.30cm}
Kyungmin Kim and Hyun Kyu Lee

\vskip 0.50cm

{ \it Department of Physics, Hanyang University, Seoul 133-791, Korea}

\end{center}
\vskip 1cm

\centerline{\bf Abstract}
\vskip 0.5cm

For a dense stellar matter, which is  electrically neutral and in
beta equilibrium, the electron chemical potential, $\mu_e$,  will
depend  nontrivially on baryonic  matter density.  It is generally
expected that as density increases, the electron chemical potential
will increase and new degrees of freedom will emerge as $\mu_e$
becomes comparable to their energy scales.  Assuming the electrical
neutrality and beta equilibrium for the stellar matter, we have
studied how the  density dependence of  lepton chemical potentials
varies for different models of  nuclear interactions that are
constrained by experiments  up to  nuclear matter density, $n_0$,
but extrapolate differently(unconstrained) beyond $n_0$ and
calculated the relative abundances of nucleons(neutron and
proton) and leptons(electron and muon) and their density
dependencies.  We find that the density dependence of the  electron
chemical potential  is strongly dependent on the structure of the
nuclear symmetry energy relevant to softness/stfness of the nuclear
matter EOS that measures the energy relevant to the neutron-proton
asymmetry. As a consequence, the
relative abundances of neutrons, protons, electrons, and muons as well as
the kaon condensation are strongly dependent on the nuclear symmetry
energy.  An intriguing result in our finding is that contrary
to the accepted lore, kaon condensation in neutron star matter,
which is considered to be the first phase transitions beyond $n_0$
and plays a crucial role in certain scenarios of compact-star
formation, is not directly tied to the softness or stiffness
of the EOS beyond $n_0$. This point is illustrated with a
''super-soft" EOS that is fit to the $\frac{\pi^-}{\pi^+}$ ratio
data of GSI which excludes kaon condensation at any density.

\end{titlepage}\vskip 0.5cm

\section{Introduction}
\setcounter{equation}{0}

\renewcommand{\thepage}{\arabic{page}}
\setcounter{page}{1} \setcounter{footnote}{0}
\renewcommand{\theequation}{\thesection.\arabic{equation}}

Recent interest on the astrophysical compact objects, i.e., neutron stars and black holes, and their evolutions(cooling and collapse for example)
brings out a very challenging question as to what is the relevant equation of state(EOS) of the
stellar matter, deep inside which the density is expected to be much  higher than the normal nuclear density,
$n_0$.    There have been lots of development in  constructing the equations of state for nuclear matter, which
are model dependent but supposed to be
consistent with the experiments up to $n_0$\cite{nuc}.
However,  the simple extrapolations of equations
of states beyond $n_0$ lead to different and strongly model-dependent results for the  macroscopic variables: chemical
potentials of constituents,  energy density, pressure, etc..  And it becomes a very important subject to
determine or predict EOS far beyond normal nuclear density.

The difference of stellar matter from the nuclear matter is that there are additional constraints for a stellar
matter, which  are
the conditions of electrical neutrality and beta equilibrium.  These constraints  naturally lead to
the emergence of leptons(electron and muon) as well as  protons    and   the lepton chemical potential is
 expected to be increasing with the density.   It opens  new degrees of freedom for the stellar matter when
the scale for the new degrees of freedom is comparable to the lepton  chemical potential.

For an ideal noninteracting fermi gas of neutron and proton the
lepton chemical potential is determined by the Fermi-Dirac
statistics only and is known to be increasing slowly with the
nucleon density. But the  nucleon system is very strongly
interacting  and  the isospin asymmetric nuclear interaction, which
is known as symmetry energy of nuclear matter, cannot be simply
ignored in the stellar matter.  The symmetry energy is defined by
the sum of the kinetic contribution and asymmetric nuclear
interactions, which are required to have neutron-proton asymmetry.
The density dependencies of the symmetry energy have been discussed
in the various theoretical models \cite{nuc}\cite{kut}\cite{ll}  and
also have been used for the phenomenological applications including
heavy ion experiment\cite{li02} and astrophysics.  Among the
interesting issues, where the importance of symmetry energy in
astrophysical phenomena has been well addressed, are the size of
neutron stars\cite{ainsworth}\cite{lattimer}\cite{piekarewicz}\cite{LS}, the
onset of kaon condensation\cite{llb97}\cite{LP}, the cooling of
proto-neutron star\cite{lpph}, gravitational waves\cite{WLK} and the instability of neutron
stars\cite{WLC}.

In this work we focus on the density dependent change of relative
abundances of the stellar matter with particular emphasis  on role
of the nuclear symmetry energy\cite{SZ}\cite{BB},  which we try to
elaborate  more  transparently  using the various models of symmetry
energy. As noted above,
as density is increasing, we are expecting the emergence of new degrees of freedom
very essential in formulating the corresponding EOS, which is only possible when we know the abundances of the constituents.
  Also we could get a more transparent understanding of EOS  in terms of
  its constituents.

The phenomenological models of the symmetry energy show  similar
density dependencies up to near the normal nuclear density but their
density dependencies diverge quite widely beyond the nuclear density from  model to
model(for a recent review see Ref. \cite{nuc}).   Therefore any simple extrapolation far beyond
normal density should result in quite different conclusions, which
might be the case for investigating stellar matter with higher
density.  In this work we calculate the density dependence of
the abundances using  a  few selective  models of nuclear
symmetry energy, which  have been used extensively to calculate neutron star EOS in detail\cite{nuc}\cite{piekarewicz}\cite{llb97}\cite{LP}\cite{tpl94}.
We demonstrate how the lepton chemical potentials
depends on the nuclear density and also how the results  depend on  models of the nuclear symmetry energy.
It is worth noting that $\pi^-/\pi^+$ ratio in heavy ion collisions  has been known to be sensitive on the symmetry energy.  Recent studies by Xiao et al. \cite{xiao} of FOPI data  at SIS/GSI seem to show that the super-soft EOS that gives the x=1 curve in \cite{xiao} is the only EOS so far available that fits the pion data of GSI.  However  the symmetry energy factor with x=1 drops to zero before reaching higher densities than $\sim 3 n_0$, which on the other hand cover the  relevant range of density in a stellar matter.
 Hence the  beta equilibrium in stellar matter might not be expected  nor the kaon condensation as discussed below.

When the density is increasing in stellar matter new degrees of
freedom, excited baryons and mesons including hyperons, are
supposed to emerge.  Kaon is among the possible new degrees of
freedoms after light leptons, when the medium
 dependent energy of kaon becomes comparable to the lepton chemical potential.
   Employing the simple formulae for the inverse propagator of kaon in medium\cite{chlee},
   the threshold densities for the kaon condensation are calculated using
   the model-dependent symmetry energies.  It is a very interesting issue since Bethe-Brown argument\cite{bethe} for the maximum mass of neutron star
,  $M^{max}_{NS} \sim 1.5 M_{\bigodot}$, depends on the onset of the kaon condensation at the density of $\sim 3n_0$.

  In this work,   the density dependence of the relative abundances of stellar
   constituents( neutron, proton, electron and muon)are calculated using
several models of density dependent nuclear symmetry energy, since it  is directly related to the relative abundances
   of constituents.
    For the possible implication
   of the kaon condensation,  the threshold densities of kaon condensation are
   calculated by comparing the lepton chemical potentials with the zero of the kaon
   inverse propagator.

Let us start with  a  simple  matter consisting of pure neutron gas.
The temperature is taken to be zero throughout this work. The energy
density and pressure of a simple free neutron gas\cite{st}(We put
$\hbar = c = 1$) are given by : \be
\epsilon&=& \frac{8 \pi}{(2\pi)^3} \int^{p_F}_0 \left(p^2 + m_n^2\right)^{1/2} p^2 dp \\
P &=& \frac{8 \pi}{3(2\pi)^3} \int^{p_F}_0 \left(p^2 +
m_n^2\right)^{-1/2} p^4 dp \ee where $p_F$ is the Fermi momentum
defined by  the neutron number density \be n = \frac{1}{3 \pi^2}
p_F^3. \ee The Fermi momentum in unit of  normal nuclear density ,
$n_0=0.16 \fm^{-3}$, and the kinetic energy(Fermi level) is given by
\be p_F = 336 \MeV \left(\frac{n}{n_0}\right)^{1/3}, ~~~~ E_F^{kin}
= \frac{p_F^2}{2 m_N} = 60 \MeV \left(\frac{n}{n_0}\right)^{2/3}.
\ee  The chemical potential of neutron, $\mu_n$, is the same as the
Fermi energy, $E_F$ : \be \mu_n = E_F \equiv \left( p_F^2 + m_n^2
\right)^{1/2} =  m\left(1+x^2\right)^{1/2}. \label{chemx} , \ee
where $x$ is a dimensionless parameter
 defined by
\be x = \frac{p_F}{m}. \ee

It is interesting to note that there is isospin symmetry in nature.
A proton is the isospin partner of the neutron, which consists of an
iso-doublet with almost the same mass, $\sim 940 \MeV$. Hence if
there is any channel for the neutron to be converted to proton, the
energy of the system of neutron only can be lowered. The conversion
depends on the dynamics, which is supposed to be the $\beta$-decay,
\be
 n \rightarrow p + e + \bar{\nu}_e
\ee
to reach $\beta$-equilibrium eventually for which
\be
\mu_n = \mu_p + \mu_e + \mu_{\nu}.
\ee Hereafter we will assume all neutrinos are emitted out of star, then we take
the neutrino chemical potential as $\mu_{\nu}=0$, $\mu_n = \mu_p + \mu_e$.

If the  energy only is concerned, then the minimum can be achieved when
$\mu_n=\mu_p$ without electron, $\mu_e =0$ .
However, a stellar matter is believed to wind up as a charge neutral object.  Practically  during the period
of reaching the $\beta$ equilibrium, the stellar matter reacts to satisfy the neutrality condition, \be n_p = n_e
\textrm{\phantom{a} or \phantom{a}} m_p x_p = m_e x_e, \label{npeneutral}\ee where
$x_p$ and $x_e$ are the dimensionless Fermi momentum of proton and of
electron, respectively. This implies the number of protons should be
balanced by the number of electrons, which means the energy minimum
can be reached with the constraint of neutrality for a system of  free neutron, proton
and electron(NPE gas) in a $\beta$ equilibrium.  It is one of the
essential differences from nuclear matter, where the charge
neutrality is not a physical constraint, for example, for  heavy ions.

As the density increases such  that the electron chemical potential
is comparable to muon mass, $m_{\mu}$, it is energetically favorable
for the electrons to convert to muons as $e \rightarrow \mu + \nu_e
+\bar{\nu}_{\mu}$. Then the chemical equilibrium can be accomplished
as \be \mu_n - \mu_p = \mu_e = \mu_{\mu}, \label{npemuchem} \ee with
the neutrality condition,   \be n_p = n_e + n_{\mu}\label{muneutr}.
\ee At the muon threshold, the muon density is zero, $n_{\mu}=0$ or
$x_{\mu}=0$ and $ \mu^{th}_{e} = m_{\mu} =106\MeV$ and the Fermi
momentum of electron at threshold is determined by the masses of
muon and electron as \be x_e |_{\mu-\textrm{thres}} =
\left[\left(\frac{m_{\mu}}{m_e}\right)^2-1\right]^{1/2} = 207. \ee
The nucleon density at threshold can be determined by Eq.
(\ref{muneutr}), $n_p = n_e$,  \be n |_{\mu-\textrm{thres}} = 2.91
n_0,  ~~~\frac{n_p}{n} |_{\mu-\textrm{thres}} =0.011.
\label{npemuonth}\ee Beyond the muon threshold the constituents of
the stellar matter becomes neutron, proton, electron and
muon(NPE$\mu$ gas).

\begin{figure}[t!]
\begin{center}
\includegraphics[height=10.0cm]{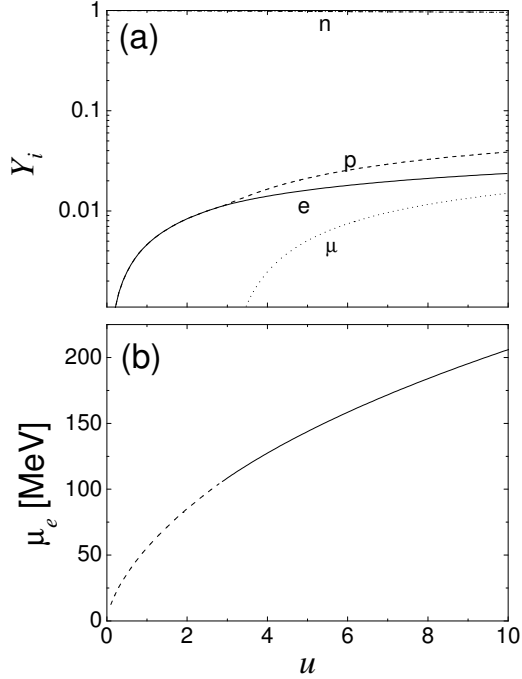}
\caption{(a)The relative  abundances of particles and (b)$\mu_e$ vs. $u$. Solid line and dashed line respectively
refer to NPE$\mu$ gas and NPE gas.}
\label{npemu-Yi-mue}
\end{center}
\end{figure}
The relative abundances of neutron, proton, electron and muons are
shown in Fig. \ref{npemu-Yi-mue}(a).  The relative neutron abundance
is almost not changing but the relative abundances of proton are
increasing substantially beyond the muon threshold as expected. The
proton fraction, $n_p/n$,  increases  with density as well as for
the electron,  $n_e/n$.

The electron chemical potential increases with density as
shown in Fig. \ref{npemu-Yi-mue}(b). In the NPE gas, the number of electrons is equal to that
of protons by the charge neutrality condition, Eq. (\ref{npeneutral}). Until the muon threshold density($\simeq 3n_0$), the increment of the electron chemical potential
follows  the property of the NPE gas. Then, beyond the threshold density, the electron
chemical potential grows with density by the property of the NPE$\mu$ gas.  In
the NPE$\mu$ gas electron shares its number with muon  such that
the number of electron is reduced by the presence of muon.

\section{Relative abundances with nuclear symmetry energy}
\setcounter{equation}{0}

So far, in calculating the chemical potentials, the nucleons are  considered  as free particles,
\be
\mu_n^{free} - \mu_p^{free} = m_n (1+x_n^2)^{1/2} - m_p(1+x_p^2)^{1/2},\label{npfree}
\ee
which can not be realistic due to the nuclear interactions.

 It is useful to write the energy per particle in the nuclear
matter \cite{p94} as \be E(n,N_p) \simeq m_N + \frac{3}{5}E_F^0
\left(\frac{n}{n_0}\right)^{2/3} + S(n)(1-2N_p)^2 + V(n), \ee where
the last term does not depend separately on neutron or proton number
density but only on total number density. In the second term, $E_F^0
= \frac{(3 \pi^2 n_0/2)^{2/3}}{2m_N}$ is the Fermi energy at,
$n=n_0/2$.  The third term is the symmetry energy for nuclear matter
\be E_{sym} = S(n)(1-2N_p)^2, \ee and  the nuclear symmetry energy
density, \be \epsilon_{sym} = n S(n) \left(1-2N_p\right)^2,
\label{sn} \ee where the symmetry energy factor, $S(n)$, is a model
dependent function of density\cite{tpl94}\cite{p94}\cite{bao-an}
\cite{centelles}, which is manufactured to be consistent with
nuclear matter data up to the nuclear matter density, $n_0$.

Now we can calculate the contributions of the symmetry energy density, $\epsilon_{sym}$, to the
chemical potentials of proton and neutron
to get
\be \mu_n - \mu_p = \mu_n^{sym} - \mu_p^{sym} =
4\left(1-2N_p\right) S(n). \label{sn} \ee
We can see that the larger $S(n)$, the lager the chemical potential difference.

For an ideal free gas, the chemical potential difference, in the
non-relativistic limit  of Eq. (\ref{npfree}) is given
by\cite{kutschera94} \be \mu_n^{free} - \mu_p^{free} =
\frac{1}{2m_N}\left(3\pi^2n\right)^{2/3}\left[\left(1-N_p\right)^{2/3}-N_p^{2/3}\right],
\ee which can be approximated in the well known form\cite{xiao}, \be
\mu_n^{free} - \mu_p^{free} = 4\left(1-2N_p\right)S_{free}(n), \ee
where \be S_{free}(n) = \left(2^{2/3}-1\right)\frac{3}{5}E_F^0
\left(\frac{n}{n_0}\right)^{2/3}. \ee There are a number of
parameterizations for the symmetry energy factor, $S(n)$, which take
account of the free kinetic contribution as well as the potential
energy contribution. One of the parametrization used in Refs.
\cite{tpl94}\cite{p94} is \be S_F(n) = (2^{2/3}-1)\frac{3}{5}E_F^0
\left[\left(\frac{n}{n_0}\right)^{2/3}-F(n)\right] + S_0 F(n),
\label{s2}\ee where  $F(n)$ is a parameter for the potential
contribution to the symmetry energy, which  satisfies $F(0)=0$ and
$F(n_0)=1$. Hereafter we set $F(n)$ simply as $F(n)=n/n_0$. In the
last term of Eq. (\ref{s2}), $S_0$ denotes the bulk symmetry energy
parameter, $S_0 \simeq 30\MeV$.

One of the different forms of $S(n)$ suggested in\cite{bao-an} is
\be
S_\alpha = (2^{2/3}-1)\frac{3}{5}E_F^0\left(\frac{n}{n_0}\right)^{2/3} + A(\alpha) \frac{n}{n_0} + \left[18.6 - A(\alpha)\right]\left(\frac{n}{n_0}\right)^{B(\alpha)}, \label{symbao-an}
\ee
where $\alpha$ is a free parameter which determines the bulk property of nuclear matter. In this work, we take $\alpha=1$ , which reproduces $\pi^+ / \pi^-$ ratio in heavy ion collision\cite{xiao}.
\be
A(\alpha=1) \simeq 107\MeV \mbox{   and   } B(\alpha=1) \simeq 1.25
\ee
(for other parameter sets, see Ref. \cite{bao-an}). In Eq. (\ref{symbao-an}), the first term denotes the kinetic contribution and last terms
refer to the potential energy contribution of nuclear matter as in $S_F$.

There is also a different type of symmetry energy factor, in which different parametrization scheme is adopted as\cite{centelles},
\be S_3(n) \simeq S^*_0 + L\rho + \frac{1}{2}K \rho^2, \label{bulkpara}
\ee
where
\be \rho = \frac{n-n^*}{3n^*},
\ee
with $n^* = 0.148 \fm^{-3}$  and $S_3$ represents either $S_{\textrm{FSU}}$
or $S_{\textrm{NL3}}$ in Ref. \cite{centelles}.
The bulk parameters, $S_0^*$, $L$ and $K$ in Eq. (\ref{bulkpara}) consistent with the nuclear matter
are listed in Table 1.

\begin{table}
\begin{center}
\begin{minipage}{.8\textwidth}
\begin{center}
\begin{tabular}{c c r@{.}l r@{.}l}
\hline
\hline
Model & $S^*_0$   & \multicolumn{2}{c}{$L$}  &   \multicolumn{2}{c}{$K$} \\
\hline
FSU   & $32.59$ & $60$ & $5$  & $-51$ & $3$ \\
NL3   & $37.29$ & $118$ & $2$ & $100$ & $9$ \\
\hline
\hline
\end{tabular}
\caption{Bulk parameters for the symmetry energy factor in Ref. \cite{centelles}.}
\end{center}
\end{minipage}
\label{S3-param}
\end{center}
\end{table}

\begin{figure}[t!]
\begin{center}
\includegraphics[height=7.0cm]{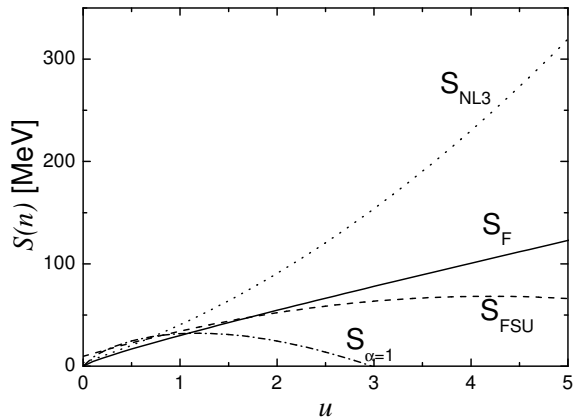}
\caption{The density dependencies of $S(n)$'s for different models.}
\label{Ss}
\end{center}
\end{figure}

For comparison, we plot $S_F$, $S_{\alpha=1}$, $S_{\textrm{FSU}}$
and $S_{\textrm{NL3}}$ in Fig. \ref{Ss}..
As one can see these symmetry energy factors are not much different from each other  up to
$n_0$. But beyond $n_0$, the difference becomes significant from
model to model as shown in Fig. \ref{Ss}. In general the symmetry energy factors are increasing with the density
up to $n \sim 5n_0$. However $S_{\alpha=1}$ has peak near $n \simeq n_0$ and drops afterwards to zero at $n \sim 3n_0$.

\begin{figure}[t!]
\begin{center}
\includegraphics[height=7.0cm]{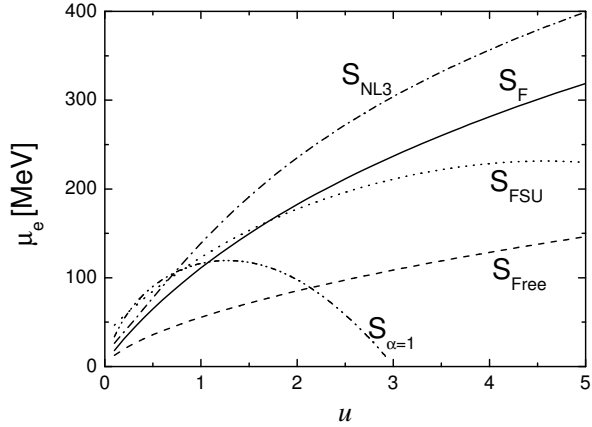}
\caption{The electron chemical potentials, $\mu_e$ due
to different models for sNPE gas.}
\label{snpe-mue}
\end{center}
\end{figure}

The electron chemical potential and its density dependence in $\beta$-equilibrium can be calculated using
Eqs. (\ref{npemuchem}) and (\ref{sn}) for different symmetry energy factors, as shown in Fig. \ref{snpe-mue}.
The electron chemical potential for the free neutron gas is lower than with other symmetry energy factors, except $S_{\alpha}$,  beyond $n \simeq  2 n_0$. This is the main reason why it is difficult to excite new degrees of freedom in the lower density  for the case of the free nucleon gas.
For $S_{\alpha=1}$, the density dependence of  the electron chemical potential is quite different such that it drops to zero $n \simeq 3n_0$.  It implies  that  $\beta$-equilibrium may not be reached for $n \gsim 3n_0$ with the symmetry energy factor of $S_{\alpha=1}$.

When  the electron chemical potential increases
such that it becomes comparable to the
rest mass of muon, muons become  constituents of stellar matter in addition to  neutrons, protons and  electrons. At the muon threshold with $n_\mu=0$ the electron chemical potential is just the mass of muon as discussed in previous section.
Then with given proton and electron densities, we can determine the nucleon density, $n |_{\mu-\textrm{thres}}$, at muon threshold from the chemical equilibrium condition, given by
\be 4(1-2N_p)S(n) =
m_e\left(1+x_e^2\right)^{1/2}. \label{smuth} \ee

In the previous section, we get $n_{\mu-\textrm{thres}}=2.91n_0$ in Eq. (\ref{npemuonth}) for free nuclear matter. With the symmetry energy factor, $S_F(n)$, Eq. (\ref{smuth}) gives
\be
n |_{\mu-\textrm{thres}} = 0.94 n_0,
\ee
which is much lower than for the free case. Threshold densities  for other symmetry energy factors can be easily guessed in Fig. \ref{snpe-mue} and the corresponding threshold densities are calculated using Eq. (\ref{smuth}) in Table 2. One can note that the muon threshold density becomes about three to  four times lower than for the free nuclear
matter.

\begin{table}
\begin{center}
\begin{minipage}{.8\textwidth}
\begin{center}
\begin{tabular}{c | c c c c c}
\hline
\hline
{} & $\mbox{Free}$ & $S_F$ & $S_{\mbox{FSU}}$ & $S_{\mbox{NL3}}$ & $S_{\alpha=1}$\\
\hline
$n_{\mu-\mbox{thres}}$ & $2.91$ & $0.94$ & $0.76$ & $0.73$ & $0.73$\\
\hline
\hline
\end{tabular}
\caption{The  muon threshold densities obtained using different
models for $S(n)$, in  unit of $n_0$.}
\end{center}
\end{minipage}
\label{mu-thres}
\end{center}
\end{table}

\begin{figure}[t!]
\begin{center}
\includegraphics[height=7.0cm]{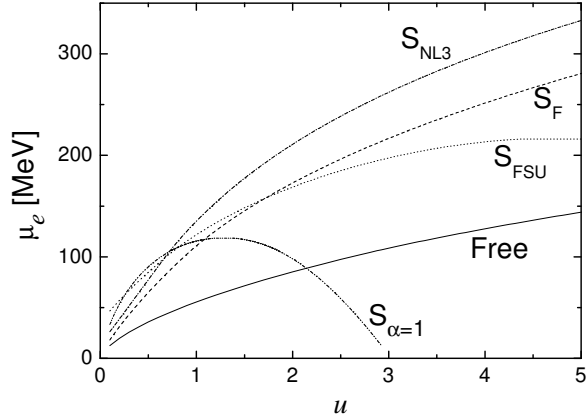}
\caption{The electron chemical potentials for sNPE$\mu$ gases with different nuclear symmetry energy models.}
\label{snpemu-mue}
\end{center}
\end{figure}

\begin{figure}[t!]
\begin{center}
\includegraphics[height=10.0cm]{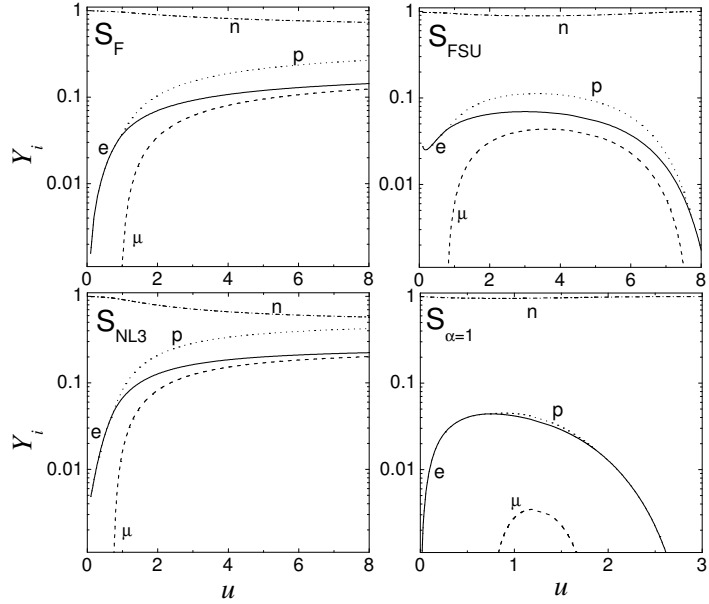}
\caption{The relative abundances of constituent particles for
sNPE$\mu$ gases with different nuclear symmetry energies.} \label{snpemu-Yis}
\end{center}
\end{figure}

At higher density beyond the muon threshold, the stellar matter
consists of neutron, proton, electron and muon(sNPE$\mu$ gas). And
the chemical equilibrium condition and charge neutrality condition
become \be
4(1-2N_p)S(n) &=& m_e(1+x_e^2)^{1/2} = m_{\mu}(1+x_{\mu}^2)^{1/2},\\
n_p &=&  n_e + n_{\mu}.
\ee
Effectively these are three equations to be solved for four unknowns, $x_e, x_{\mu}, x_n$ and $x_p$.
Then for a given electron chemical potential, $x_e$, one can solve these equations completely to determine the chemical potentials and the relative abundances of the constituents, $n_e, n_{\mu}, n_n$, and $n_p$.  The electron chemical potentials(same as the muon chemical potentials in beta equilibrium) beyond the muon threshold densities(see Table 2) are calculated for different models of the nuclear symmetry energy as plotted in Fig. \ref{snpemu-mue}.  The density dependence of electron chemical potential  for $S_{FSU}$  shows a quite different  behavior  from others.   One can observe in Fig. \ref{snpemu-Yis} that the proton  and muon fractions are increasing with the density much faster than in free case.  For $S_{FSU}$, the fractions of proton and muon are increasing with the density up to $\sim 4 n_0$ but decreasing and even becomes lower than free case for the higher density. The $S_{\alpha}$ case shows a similar behavior with $S_{FSU}$, but the lepton and proton fractions are decreasing more quickly than $S_{FSU}$ and the muon and proton fractions, beyond the muon threshold density, are much lower than in other models.

\begin{figure}[t!]
\begin{center}
\includegraphics[height=10.0cm]{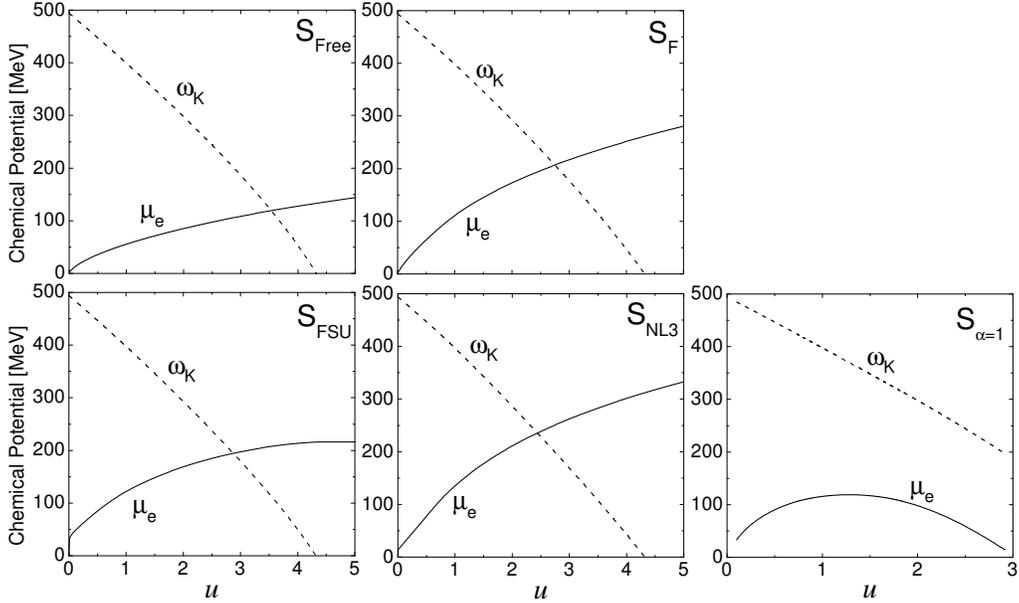}
\caption{The electron chemical potentials and kaon chemical potentials for different nuclear symmetry energy models.}
\label{kaon-chem}
\end{center}
\end{figure}

 When the leptonic chemical potentials  increase high enough, it is natural to ask what kind of new degrees of freedom
 can be driven to evolve in a stellar matter.  They could be excited baryons including hyperons as well
 as strange mesons. One of the interesting possibilities is the s-wave kaon condensation\cite{brown}
 driven by the electron chemical potential.  The simplest way of demonstrating the possibility of kaon condensation
 is by comparing the electron chemical potential with the kaon chemical potential at the threshold, which determines
 the kaon threshold density.  At the threshold the kaon chemical potential is given by
 the zero of the inverse propagator, $D_{K^{-}}^{-1}$, with zero momentum\cite{chlee} which is given  by

 \be
D_{K^{-}}^{-1} &=&   \omega_K^2 - m_{K}^2 +\frac{1}{f^2}(n_n/2 + n_p) \omega_K +  \frac{\Sigma_{KN}}{f^2}n, \label{muK}
\ee
where $f$ is the pion decay constant, $f=93$MeV. The possible range of
$\Sigma_{KN}$\cite{chlee}\cite{lbmr95}\cite{chlee96}\cite{liu} is estimated to be
 \be
    \Sigma_{KN} = 200 - 400 \MeV.
 \ee
As one of the example, we take  $\Sigma_{KN} = 400 \MeV$  in this work\cite{llb97}.  It is found that  up to
 $n \sim 5n_0$ the dependence of $\omega_K(n)$ on the symmetry energy models is not
so significant and the kaon threshold densities are estimated around $n \sim  3 n_0$ as shown in Table 3 and
in Fig. \ref{kaon-chem}.   The threshold density is lower $ \sim 2.5n_0$ with $S_{NL3}$ but  higher for free case,
 $\sim 3.5 n_0$.  One should note that  $\omega_K$ obtained using Eq. (\ref{muK}) is found to be always larger than
 electron chemical potential such that  one cannot expect a   kaon threshold if   $S_{\alpha=1}$ is valid up to
  $n \sim 3n_0$.

\begin{table}
\begin{center}
\begin{minipage}{.8\textwidth}
\begin{center}
\begin{tabular}{l | c c c c}
\hline
\hline
 & $\mbox{Free}$ & $S_F$ & $S_{FSU}$ & $S_{NL3}$ \\
\hline
 $n \left[n_0\right]$ & $3.53$ & $2.76$ & $2.89$ & $2.45$\\
\hline
 $\mu \left[\MeV\right]$ & $120.4$ & $206.3$ & $192.5$ & $234.5$\\
\hline
\hline
\end{tabular}
\caption{The  kaon threshold densities estimated using different models for $S(n)$ and
the electron chemical potentials at the threshold densities.}

\end{center}
\end{minipage}
\label{kaon-thres}
\end{center}
\end{table}

\section{Discussion}

In this work we consider the effect of the nuclear symmetry energy on the relative abundances of particles,
neutron, proton, electron and muon, with the nucleon density inside a stellar matter supposed to be
electrically neutral and in beta equilibrium. We determine the electron and muon threshold densities.
The muon threshold density  reduces substantially with symmetry energy from that of free nucleon gas.
It is observed that  the  relative abundances    with  symmetry energy  diverge significantly from model
 to model for higher density $n>n_0$.  We also estimate the kaon condensation threshold densities using a
 number of different models for the symmetry energy.  The kaon threshold estimated with symmetry energy
  considered is found to be lower than the free case. However, for the super-soft EOS, in which  the symmetry energy
  factor $S_{\alpha=1}$ drops to zero  before reaching higher density than $\sim 3 n_0$,
   beta equilibrium in stellar matter cannot be expected.  Moreover   $\omega_K$ obtained using Eq.(\ref{muK})
    is found to be always larger than electron chemical potential such that  one can not find  kaon threshold
     for   $S_{\alpha=1}$. This is an intriguing consequence of our finding
     that contrary to the  accepted lore, kaon condensation in
     neutron-star matter which is considered to be the first phase
     transition as density increase beyond $n_0$ and hence plays a
     crucial role in certain scenarios of compact-star formation is
     indifferent to the softness/stiffness of EOS beyond $n_0$.

 We can notice that  the presently known models for the symmetry energy factor provide quite different
 predictions  for higher densities when the parametrization is straightforwardly extrapolated to higher
  densities beyond $\sim 3 n_0$,  which are however  very important deep inside the stellar system.
  Hence the predictions on the stellar matter structure after integrating the Tolman-Oppenheimer-Volkov
  equation should depend strongly on the specific form of
  symmetry energy\cite{lattimer}\cite{piekarewicz}\cite{LP}\cite{steiner}.
  Hence  it is very important to find the symmetry energy with correct density dependency for a stellar
   matter and we are  looking forward to get  more experimental information on symmetry energy from the
   forthcoming experiment like FAIR/GSI.

One of the interesting ideas developed recently  in hadron
physics is adopting the hidden local symmetry\cite{harada}, which
opens up an idea of hadronic freedom\cite{brownrho}. However it is not
well investigated how  this idea can be employed in understanding
the symmetry energy at higher density and its relation to the
softness/stifness of EOS. Whether it simply converges to the free
case for higher density than $\sim 3 n_0$ or it is quite different
matter from the free fermi gas and how the symmetry energy is
related to the KN interaction which controls  the kaon condensation
at high density nuclear matter are among the very intriguing
questions to be answered.     Since it is the region so far
unexplored by the experiment, it is very interesting to construct a
form of symmetry energy compatible with the new idea of dense
hadronic matter  to investigate  the implications on the stellar
structure using TOV equation, which will be discussed in a separate
work.

\subsection*{Acknowledgments}
We are grateful to  Chang-Hwan Lee, Bao-An Li and Mannque Rho for the very helpful discussions.
This work is supported by the WCU project of Korean Ministry of Education, Science and Technology
(R33-2008-000-10087-0).  KK is also supported by the Seoul Fellowship of Seoul Metropolitan Government.

{}


\begin{thebibliography}{}

\bibitem{nuc} B.-A. Li, L.-W. Chen and C.M. Ko, Phys. Rept. 464, 113 (2008) and references therein.


\bibitem{kut} M. Kutschera, Z. Phys. A 348, 263(1994)

\bibitem{ll} C.-H. Lee, T.T.S. Kuo, G.Q. Li, and G.E. Brown, Phys.
Rev. C 57, 3488(1998)

\bibitem{li02} B.-A. Li, Phys. Rev. Lett. 88, 192701(2002)


\bibitem{ainsworth} M. Prakash and T. Ainsworth, Phys. Rev. C 36, 346 (1987).

\bibitem{lattimer} J.M. Lattimer, Lecture note of the Connecting Quarks with the Cosmos Summer School,
Seattle, Washington, U.S.A., 2009 (unpublished).

\bibitem{piekarewicz} J. Piekarewicz, AIP Conf. Proc., 1128, 114 (2009).

\bibitem{LS} B.-A. Li and A.W. Steiner, Phys. Lett. B642, 436(2006)


\bibitem{llb97} G.Q. Li, C.-H. Lee and G.E. Brown, Nucl. Phys. A625, 372 (1997).

\bibitem{LP} J.M. Lattimer and M. Prakash, Phys. Rep. 333, 121 (2000); Astrophys. J. 550, 426 (2001);
 Phys. Rept. 442, 109 (2007).

\bibitem{lpph} J.M. Lattimer, C.J. Pethick, M. Prakash, P. Haensel,
Phys. Rev. Lett. 66 (1991) 2701

\bibitem{WLK} D.-H. Wen, B.-A. Li and P.G. Krastev, [arXiv:0902.4702].

\bibitem{WLC} D.-H. Wen, B.-A. Li and Lie-Wen Chen, arXiv:0908.1922

\bibitem{SZ} K. Sumiyoshi, H. Suzuki and H. Toki, Astron. Astrophys. 303, 475(1995)

\bibitem{BB} S. Banik and  D. Bandyopadhyay, J. Phys. G 26, 1495(2000)

\bibitem{tpl94} V. Thorsson, M. Prakash and J.M. Lattimer, Nucl. Phys. A572, 693 (1994).

\bibitem{xiao} Z. Xiao, B.-A. Li, L.-W. Chen, G.-C. Yong and M. Zhang, Phys. Rev. Lett. 102, 062502 (2009).

\bibitem{chlee} G.E. Brown, C.-H. Lee and M. Rho, Phys. Rept. 462, 1 (2008).

\bibitem{bethe} G.E. Brown and H.A. Bethe,  Astrophys. J. 423, 659 (1994).

\bibitem{st} S. Shapiro and S. Teukolsky, \textit{Black Holes, White Dwarfs, and Neutron Stars:
The Physics of Compact Objects}
(Wiley-VCH, New York, 2004).

\bibitem{p94} M. Prakash, Proceeding of the Nuclear Equation of State, Puri, India, 1994, edited by A. Ansari and L.S. Satpathy (World Scientific, Singapore, 1996), p. 229-410.

\bibitem{bao-an} C.B. Das, S. Das Gupta, C. Gale and B.-A. Li, Phys. Rev. C 67, 034611 (2003); A.W. Steiner and B.-A. Li, Phys. Rev. C 72, 041601(R) (2005); L.-W. Chen, C.M. Ko and B.-A. Li, Phys. Rev. Lett. 94, 032701 (2005).

\bibitem{centelles} J. Piekarewicz and M. Centelles, Phys. Rev. C 79, 054311 (2009).

\bibitem{kutschera94} M. Kutschera, Phys. Lett. B. 340, 1 (1994).

\bibitem{brown} D.B. Kaplan and A.E. Nelson, Phys. Lett, B175, 57(1986); G.E. Brown, K. Kubodera and M. Rho, Phys. Lett. B192, 273(1987).

\bibitem{lbmr95} C.-H. Lee, G.E. Brown, D.P. Min and M. Rho, Nucl. Phys. A585, 401 (1995).

\bibitem{chlee96} C.-H. Lee, Phys. Rept. 275, 255 (1996).

\bibitem{liu} S.J. Dong, J.-F. Laga\"{e} and K.F. Liu, Phys. Rev. D 54, 5496 (1996).

\bibitem{steiner} B.-A. Li and A.W. Steiner, Phys. Lett. B 642, 436 (2006).

\bibitem{harada} M. Harada and K. Yamawaki, Phys. Rept. 381, 1 (2003).

\bibitem{brownrho} G.E. Brown, C.-H. Lee and M. Rho, Phys. Rev. C 74, 024906 (2006).

\end{thebibliography}
\end{document}